\RequirePackage{lineno}
\documentclass[twocolumn,showpacs,preprintnumbers,amsmath,amssymb]{revtex4}

\usepackage{graphicx}
\usepackage{dcolumn}
\usepackage{bm}
\usepackage{amssymb}

\def\bea{\begin{eqnarray}}
\def\eea{\end{eqnarray}}

\def\pp{\mbox{$p$-$p$} }
\def\auau{\mbox{Au-Au} }
\def\aa{\mbox{A-A} }
\def\nn{\mbox{N-N} }

\def\ee{\mbox{$e^+$-$e^-$} }
\def\ppbar{\mbox{$p$-$\bar p$} }

\def\pt{$p_t$ }

\begin{document} 
\setpagewiselinenumbers
\modulolinenumbers[5]

\preprint{Version 2.2}

\title{Parton fragment yields derived from minimum-bias jet angular correlations}

\author{Thomas A. Trainor and David T. Kettler}
\address{CENPA 354290, University of Washington, Seattle, WA 98195}


\date{\today}

\begin{abstract}

Spectrum hard components from 200 GeV \auau collisions are accurately described by pQCD parton fragment distributions, indicating that a substantial population of parton fragments is present in hadron spectra at low $p_t$. Minimum-bias angular correlations contain jet-like correlation structure with most-probable hadron momentum 1 GeV/c. In this study we convert minimum-bias jet-like angular correlations to single-particle yields and compare them with spectrum hard components. We find that jet-like correlations in central 200 GeV \auau collisions correspond quantitatively to pQCD predictions, and the jet-correlated hadron yield comprises one third of the final state.

\end{abstract}


\maketitle

 \section{Introduction}

The hydrodynamic (hydro) model has been extensively applied to heavy ion data from the relativistic heavy ion collider (RHIC)~\cite{hydrotheory,heinz,rom,bwfit}. Results are interpreted to conclude that a dense QCD medium with small viscosity nearly opaque to most partons, a strongly-coupled quark-gluon plasma (sQGP) or ``perfect liquid,'' is formed in more-central Au-Au collisions~\cite{perfect1,perfect2}. Analysis methods tend to favor those conclusions:  Methods directed toward a bulk medium tend to misinterpret low-$p_t$ features of parton fragmentation. High-$p_t$ jet analysis methods tend to disregard fragmentation structure at smaller $p_t$~\cite{hardspec}. 

Alternative analysis methods provide contrasting evidence. 
Two-component analysis of single-particle hadron spectra reveals a spectrum hard component~\cite{ppprd,hardspec} consistent with a parton fragment distribution described by pQCD~\cite{fragevo} which can masquerade as ``radial flow'' in some hydro-motivated analysis~\cite{nohydro}. Minimum-bias angular correlations suggest that a large number of back-to-back jets from initial-state scattered partons with energies as low as 3 GeV survive as jet-like hadron correlations even in central Au-Au collisions, implying near transparency to partons~\cite{axialci,hijscale,ptscale,ptedep,daugherity}. 

The goal of the present analysis is to establish a direct connection between minimum-bias jet angular correlations and spectrum hard components. 
We test the hypothesis that a jet-like ``same-side'' 2D peak at the origin in angular correlations represents all minimum-bias parton fragments which also appear as the spectrum hard component. Measured spectrum hard components from p-p and Au-Au collisions have been described by pQCD~\cite{fragevo}. By establishing a quantitative relation between jet-like correlations and spectrum yields we relate minimum-bias jet (minijet) correlations directly to pQCD theory.

A significant consequence of this analysis is confirmation that a substantial fraction of the final state in central Au-Au collisions consists of resolved jets with energies as low as 3 GeV. The evolution of nuclear collisions is apparently dominated by parton scattering and fragmentation even in the most central \auau collisions, albeit the fragmentation process is strongly modified. In this analysis we consider $p_t$-integral yields. In a follow-up analysis we will present fragment $p_t$ spectra inferred from jet angular correlations obtained within specific $p_t$ cuts.


This article is structured as follows: We introduce  parton fragment distributions as jet correlations and spectrum hard components. We describe an analysis method to convert jet correlations to hard-component hadron yields. We review measured minimum-bias jet correlations. We infer jet properties from jet correlations which provide new information on fragmentation at smaller $p_t$. We combine jet properties inferred from correlations to predict jet fragment yields. Finally, we compare our predictions with measured spectrum hard-component yields.

\section{Parton fragment distributions}


The quantitative correspondence between spectrum hard components~\cite{hardspec} and pQCD fragment distributions established in Ref.~\cite{fragevo} strongly suggests that the parton fragment interpretation for hard components is valid. However, no such correspondence has been established for minimum-bias angular correlations (without $p_t$ ``trigger/associated'' cuts). Interpretation of low-$p_t$ jet-like features as true jet structure~\cite{porter1,porter2,axialci,daugherity} is questioned by some as outside the scope of high-$p_t$ hard  processes nominally described by pQCD. Coupling spectrum hard components {\em and} minimum-bias jet-like correlations to pQCD predictions within a single quantitative system would provide compelling support for a comprehensive interpretation in terms of parton fragmentation to minijets. To achieve the connection we transform two-particle jet correlations to the equivalent single-particle fragment distributions by pair factorization.

\subsection{The two-component model}

The context of this analysis is a two-component model of fragmentation---longitudinal and transverse---manifested in 1D \pt  spectra and 2D two-particle angular correlations~\cite{ppprd,hardspec,fragevo}. The two-component model as invoked here represents two orthogonal fragmentation systems: longitudinal projectile-nucleon fragmentation (soft component) and transverse large-angle-scattered parton fragmentation (hard component). That picture is consistent with the PYTHIA Monte Carlo model of \pp collisions~\cite{pythia}. The terminology ``soft'' and ``hard'' refers to the initial \nn or parton-parton momentum transfer, not to the hadron fragment $p_t$ which may extend down to zero momentum. In this analysis we focus on fragmentation of large-angle-scattered partons manifesting as hard components in $p_t$ spectra  and jets in angular correlations. 

\subsection{Spectrum hard components}

The spectrum hard component (HC) was discovered in a \pp spectrum analysis~\cite{ppprd}. The HC was later interpreted as hadron fragments from minimum-bias jets. The HC is defined as the measured $p_t/y_t$ spectrum minus a fixed soft component (SC). The SC (interpreted as longitudinal projectile-nucleon fragmentation) is in turn defined as the limiting case of spectrum variation with event multiplicity ($p\text{-}p$) or centrality (A-A) which limit should correspond to no transversely-scattered parton fragmentation~\cite{ppprd,hardspec}. Two-component analysis of \auau collisions revealed that the HC undergoes strong evolution: with increasing centrality suppression at larger $p_t \sim 10$ GeV/c is closely correlated with much larger enhancement at smaller $p_t \sim 0.5$ GeV/c~\cite{hardspec}.

\subsection{pQCD fragment distributions}

In-vacuum \ee fragmentation functions (FFs) for a large ensemble of kinematic conditions have been measured at LEP and HERA. FFs for light quarks and gluons fragmenting to unidentified hadrons have been accurately parametrized over a large kinematic domain down to zero fragment momentum~\cite{ffprd}. A parametrization of so-called ``medium-modified'' FFs (mFFs) has also been developed~\cite{fragevo}. 
By folding a pQCD parton spectrum with parametrized \ee or \ppbar FFs or calculated mFFs we obtain fragment distributions (FDs) which provide a good description of spectrum hard components down to small fragment momentum ($p_t \sim 0.3$ GeV/c)~\cite{fragevo}.

\subsection{Minimum-bias jet angular correlations}

Minimum-bias two-particle correlations from 200 GeV \pp collisions first revealed jet-like structure (same-side 2D peak at the origin, away-side 1D ridge) as the dominant correlation structure down to small \pt ($\sim 0.35$ GeV/c)~\cite{porter1,porter2,axialci}. The jet-like structure in \pp correlations is reproduced well by PYTHIA and does not appear in PYTHIA simulations if hard scattering is disabled~\cite{porter2}. Jets individually identified in event-wise analysis of energy (calorimeter) angular distributions on $(\eta,\phi)$ follow the expected pQCD parton spectrum down to 5 GeV (3-4 GeV without background)~\cite{ua1,starpp}. Trends for jet-like correlations and event-wise reconstructed jets are consistent: The most-probable hadron $p_t$ for minimum-bias jet-like correlations in \pp collisions is 1 GeV/c, in agreement with the observed 3-4 GeV  lower limit to calorimeter-based reconstructed-jet energy spectra.

\section{Analysis method}

$p_t$-integrated jet angular correlations on 4D angle space represented by pair density $J^2(b)$ are converted to single-particle density $J(b)$ via calculated pQCD ``jet frequency'' $f(b)$. $J(b)$ is in turn compared with spectrum hard component $H_{AA}(b)$, and therefore with pQCD calculations as in Ref.~\cite{fragevo}. 

\subsection{Single-particle angular densities}

To convert jet angular correlations to fragment yields we require a model of the single-particle angular density. We denote the charged-particle 2D density on $(\eta,\phi)$ by $\rho_0 = dn_{ch}/2\pi d\eta$, with the two-component spectrum model (first line)~\cite{nardi}
\bea \label{rho0b}
 \frac{2}{n_{part}}\,  \rho_0(b)  
&=& S_{NN} + \nu\, H_{AA}(b) \\ \nonumber
&\approx&   \rho_{pp}\, \left\{1 + x\,(\nu-1) \right\},
\eea
where $\nu = 2n_{bin} / n_{part}$ is the mean participant-nucleon path length, 
$S_{NN}$ is the soft component, by hypothesis independent of \aa centrality, and $H_{AA}(b)$ is the hard component. Eq.~(\ref{rho0b}) (second line) is an alternative two-component model proposed in Ref.~\cite{nardi}. Parameter $x$ represents the (fixed) hard-component fraction of particle production. The 2D charged-particle density for $\sqrt{s_{NN}} = 200$ GeV NSD \pp (N-N) collisions is assumed in this analysis to be  $\rho_{pp} = 2.5 / 2\pi \approx 0.4$. The alternative (KN) model is further discussed in App.~\ref{knn}.




\subsection{Two-particle angular correlations}

2D angular autocorrelations are measured on difference variables $(\eta_\Delta,\phi_\Delta)$, e.g.\ $\eta_\Delta = \eta_1 - \eta_2$, within some angular acceptance $(\Delta \eta,\Delta \phi)$~\cite{inverse,axialci,daugherity}. Model fits to 2D angular correlations provide accurate separation of jet structure from nonjet structure, mainly the {\em azimuth quadrupole}~\cite{kettler} conventionally interpreted as elliptic flow~\cite{starv2}. Alternate interpretations are possible~\cite{gluequad}.

{\em Per-pair} correlations are denoted in this article by pair ratio ${\Delta \rho}/{\rho_{ref}}$, where correlated-pair density $\Delta \rho = \rho - \rho_{ref}$, $\rho$ is the mean density of {\em sibling} particle pairs from single events, and $\rho_{ref}$ is a mixed-pair reference density derived from pairs of similar events. The corresponding  {\em per-particle} correlation measure is denoted by ${\Delta \rho}/\sqrt{\rho_{ref}}$, since reference pair density $\rho_{ref}$ is approximately the product of two single-particle densities.

Model fits to 2D angular correlations isolate the same-side (SS) jet peak, approximated by a 2D Gaussian, from other structure (away-side ridge representing back-to-back jet pairs, azimuth quadrupole)~\cite{axialci,daugherity,kettler}. The SS peak pair density denoted by $J^2(\eta_\Delta,\phi_\Delta,b)$ represents the event-wise average of all {\em intrajet} correlated hadron pairs (parton fragment pairs) within the detector angular acceptance. $j^2(\eta_\Delta,\phi_\Delta,b)$ is the corresponding pair ratio in the form ${\Delta \rho}/{\rho_{ref}}$, with  $\rho_{ref} \approx \rho_0^2(b)$. The goal of the present analysis is to convert the per-event jet-correlated pair number to a parton fragment yield corresponding to the hard component of final-state hadron production.

\subsection{Parton fragment densities from jet correlations}

To convert jet-correlated pairs to single-particle fragment yields we must factorize the pair distribution. The first step is to average the SS 2D peak over the angular acceptance according to Eq.~(\ref{average}) to obtain 4D density $J^2(b) = \rho_0^2(b)\,j^2(b)$. 
Schematically,
\bea
\frac{\Delta \rho}{\rho_{ref}}(\eta_\Delta,\phi_\Delta,b) \stackrel{\text{model fit}}{\rightarrow} j^2(\eta_\Delta,\phi_\Delta,b) \stackrel{\text{average}}{\rightarrow} j^2(b).
\eea
Combining measured pair ratio $j^2(b)$ with measured single-particle 2D angular density $\rho_0(b)$ we obtain the mean two-particle 4D angular density of fragment pairs
\bea \label{bigj2b}
J^2(b) &=&  n_j(b)\, \left( \frac{n_{ch,j}}{2\pi \Delta \eta} \right)^2,
\eea
where $n_j(b)$ is the calculated per-event jet number in acceptance $\Delta \eta$, and $n_{ch,j}(b)$ is the jet fragment multiplicity. That expression represents the hypothesis that the SS 2D peak includes all intrajet correlated pairs.



Eq.~(\ref{bigj2b}) leads to a factorization assumption: The mean number of correlated hadron pairs in a jet is approximately the square of the mean jet fragment multiplicity (significant fluctuation effects are discussed in App.~\ref{flucts}). Given jet-number hypothesis $n_j(b)$ within acceptance $\Delta \eta$ the 2D fragment density on $(\eta,\phi)$ is then
\bea \label{jb}
J(b) &\equiv& n_j(b)\,  \sqrt{J^2(b)/ n_j(b)} \\ \nonumber
&=& n_j(b)\, \rho_0(b)  \sqrt{j^2(b)/ n_j(b)} \\ \nonumber
&=&n_j(b)\, \frac{n_{ch,j}(b)}{2\pi \Delta \eta}.    
\eea
$J(b)$ is the 2D angular density of parton fragments within the acceptance inferred from jet angular correlations.

\subsection{Jet correlations and spectrum hard component}

Jet correlations are related through $J(b)$ to spectrum hard component $H_{AA}(b)$ by
\bea \label{jbhaa}
\left(\frac{2\pi}{n_{bin}}\right) J(b) \hspace{-.05in} &=& \hspace{-.05in} f(b)\, n_{ch,j}(b) = \frac{dn_h}{d\eta} \equiv  2\pi H_{AA}(b),
\eea
where $(1/n_{bin})\, dn_j(b)/d\eta \equiv  f(b)$ is the jet frequency defined in Ref.~\cite{ppprd} for NSD \pp (N-N) collisions, and $dn_h/d\eta$ is the hard-component multiplicity density on $\eta$ (see Sec.~\ref{fragyield}).
Equation~(\ref{jb}) (third line) is consistent with Eq.~(9) of Ref.~\cite{fragevo}. 
Combined with Eq.~(\ref{jbhaa}) it relates measured jet angular correlations directly to hard-component yield $H_{AA}(b)$ \{integral of spectrum hard component $H_{AA}(y_t,b)$~\cite{hardspec}\} assuming that $H_{AA}$ represents a jet fragment density~\cite{fragevo}.

\subsection{Jet acceptance effects}

There are two acceptance issues for jets: (i)  fragment losses from jets at $\eta$ acceptance boundaries and (ii) the number of jets per dijet (fraction of jets with a partner) within $\eta$ acceptance $\Delta \eta$. 
The first depends on jet size relative to $\eta$ acceptance ($\sigma_\eta / \Delta \eta$), the second on fractional $\eta$ acceptance relative to $4\pi$ ($\Delta \eta/\Delta \eta_{4\pi}$).
Jet fragment multiplicities $n_{ch,j}$ are reduced from a pQCD ideal according to Eq.~(\ref{average}). The reduced multiplicity is then assumed to be point-like at the jet (parton) location on $(\eta,\phi)$.  Ideal ($4\pi$) values are indicated in relevant plots.
The number of jets in acceptance $\Delta \eta$ per dijet appearing in the full $4\pi$ acceptance is denoted by $\epsilon_j \in [1,2]$.

\section{Jet Angular correlations}

In this analysis we consider $p_t$-integral correlations as opposed to $p_t$-differential correlations on the full $(p_{t1},p_{t2})$ space or a subspace defined by $p_t$ cuts (so-called ``triggered'' correlations). 
All intra-jet hadron pairs for all jets that survive partonic and hadronic rescattering should appear in the SS 2D peak for these ``untriggered'' 
angular correlations. 

\subsection{Same-side 2D peak parametrization }

\begin{figure}[h]
\includegraphics[width=.22\textwidth]{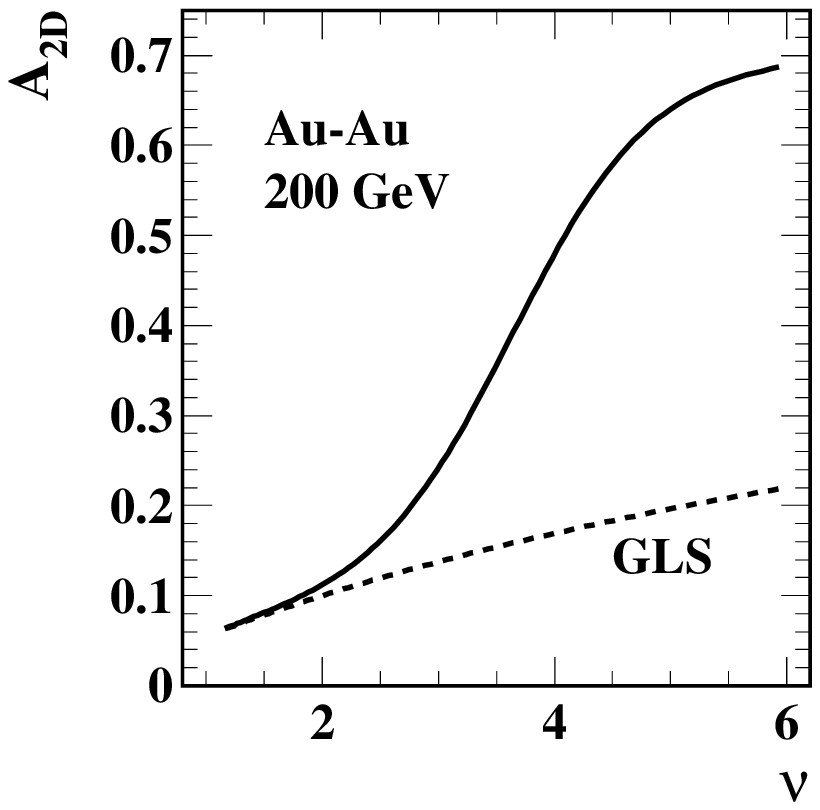}
\includegraphics[width=.22\textwidth]{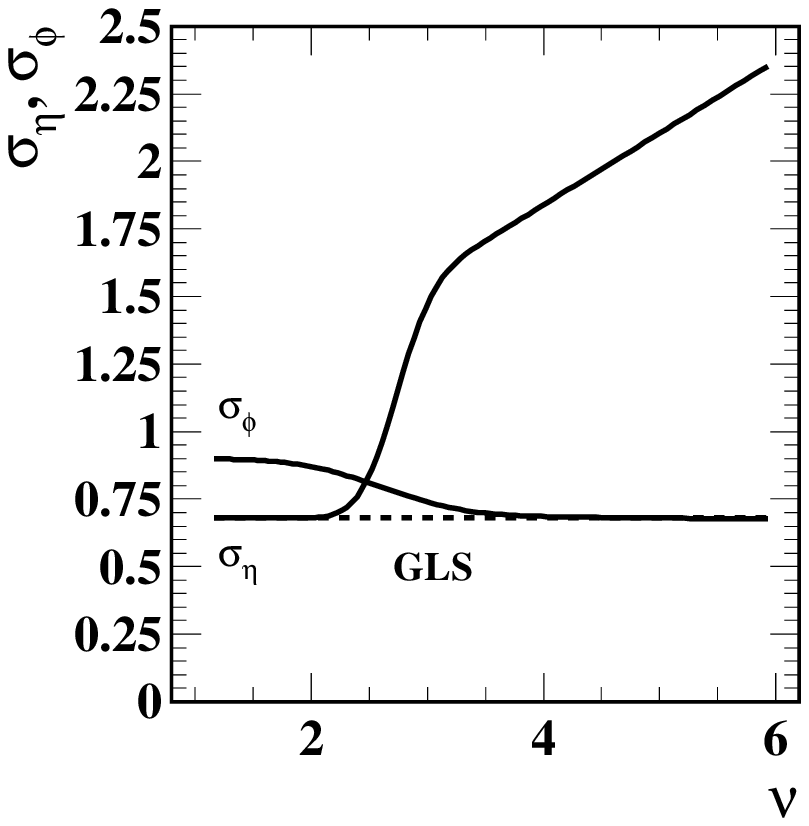}
\caption{\label{peakparams}
Fit parameters versus centrality measure $\nu$ for the same-side 2D peak in correlation data on $(\eta_{\Delta},\phi_{\Delta})$ from Au-Au collisions at $\sqrt{s_{NN}}= 200$ GeV~\cite{daugherity}. Left: The same-side 2D Gaussian amplitude. Right:  The 2D peak rms widths. 
} 
\end{figure}

Model fits to 2D angular correlations include a {same-side} (SS) jet peak modeled by a 2D Gaussian. Angular correlations in Ref.~\cite{daugherity} are reported in the {per-particle} form $\Delta \rho/\sqrt{\rho_{ref}}$. The SS 2D peak model is then
\bea \label{estruct}
\frac{\Delta \rho_{SS}}{{\sqrt{\rho_{ref}}}} &=& \rho_0(b)\, j^2(\eta_\Delta,\phi_\Delta,b) \\ \nonumber
 &=&  A_{2D}\, \exp\left(  -\eta^2_\Delta / 2\sigma^2_{\eta} \right)\, \exp\left(  -\phi^2_\Delta / 2\sigma^2_{\phi}\right).
\eea
Figure~\ref{peakparams} shows smooth parametrizations of data from Ref.~\cite{daugherity} for the SS 2D peak amplitude, $\eta$ and $\phi$ widths from $\sqrt{s_{NN}} = 200$ GeV \auau collisions. The data are described to a few percent except for the interval $\nu > 5$ where $A_{2D}$ and $\sigma_\eta$ data typically fall below the parametrizations. The systematic uncertainty in $A_{2D}$ and $\sigma_\eta$ in that region increases to 10\%. 

In more-peripheral collisions the amplitude and $\eta$ width follow {\em Glauber linear superposition} (GLS) trends (dashed curves). At a particular centrality ($\nu \sim 2.5$) jet characteristics change dramatically, with large increases in amplitude (fragment pair yield) and $\eta$ width (jet $\eta$ elongation) relative to the GLS reference.

\subsection{Angle-averaged pair ratio $\bf j^2(b)$}

\begin{figure}[h]
\includegraphics[height=1.65in]{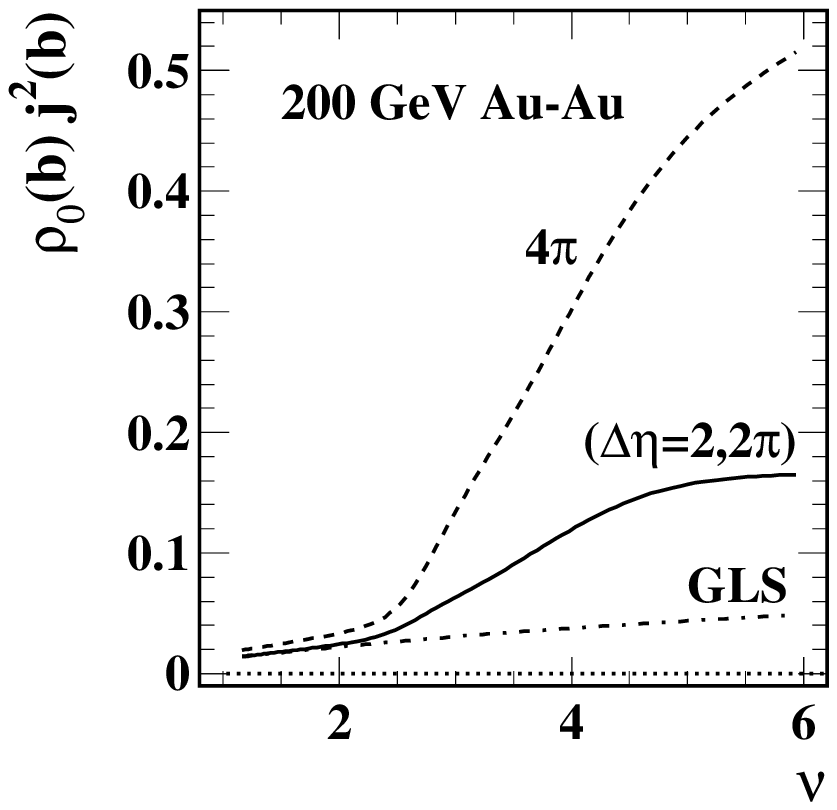}
\includegraphics[height=1.67in]{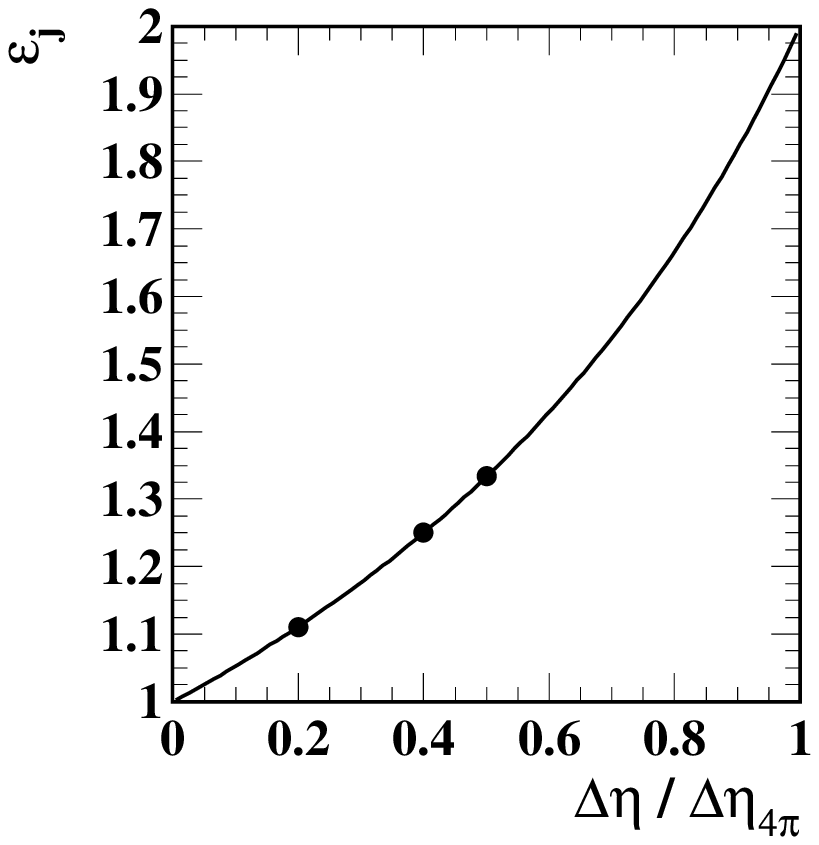}
\caption{\label{j2b}
Left: The product of single-particle 2D angular density $\rho_0(b)$ and angle-averaged pair ratio $j^2(b)$ representing the same-side 2D peak in jet angular correlations. Right: The number of jets per dijet $\epsilon_j$ within acceptance $\Delta \eta$ as a function of the $\eta$ acceptance relative to an effective $4\pi$ acceptance.
} 
\end{figure}

Figure~\ref{j2b} (left panel) shows the angle-averaged SS peak pair yield (solid curve) in the form $\rho_0(b)\, j^2(b)$ for pairs within the angular acceptance and the yield extrapolated to $4\pi$ (dashed curve).  Angle average $j^2(b)$ is defined in terms of SS peak parameters by Eq.~(\ref{average}). The dash-dotted curve is the GLS reference. The difference  between the $4\pi$ and $(\Delta \eta = 2,2\pi)$ curves is large for this pair measure because the pair acceptance depends on the square of the relative $\eta$ acceptance $\Delta \eta / \Delta \eta_{4\pi}$. The curve labeled $4\pi$ is related to the SS 2D peak ``volume'' $V$ in Ref.~\cite{daugherity} by $V/4\pi$.

\section{Jet Properties from Correlations}

We combine measured jet correlations in the form $j^2(b)$ with a pQCD estimate of jet production to determine jet frequency $f(b)$ and mean jet multiplicity $n_{ch,j}(b)$.  By removing $n_{bin}$ from jet systematics in \aa collisions we obtain two slowly-varying factors which can be studied in detail.

\subsection{pQCD fragment distributions}

A parton  {fragment distribution} (FD) on rapidity $y$ can be defined for individual NSD N-N collisions by the pQCD folding integral~\cite{fragevo}
\bea \label{fold}
\hspace{-.1in}  \frac{d^2n_{h}}{dy\, d\eta}  \hspace{-.07in} &=&  \hspace{-.07in}  \frac{\epsilon_j(\Delta \eta)/2}{\sigma_{NSD}\, \Delta \eta_{4\pi}}  \hspace{-.0in} \int_0^\infty \hspace{-.13in}  dy_{max}\, D_\text{xx}(y,y_{max},b)\, \frac{d\sigma_{dijet}}{dy_{max}} \nonumber \\
&=&   \frac{f(b)}{2\sigma_{dijet}}   \int_0^\infty   dy_{max}\, D_\text{xx}(y,y_{max},b)\, \frac{d\sigma_{dijet}}{dy_{max}} \nonumber \\
&=& f(b)\, D_{\text{xx}}(y,b)/2 = \frac{2\pi}{n_{bin}} J(y,b).
\eea

$D_{xx}(y,y_{max},b)$ is an FF ensemble from collision system xx which may therefore include ``in-medium'' modifications in \pp collisions or in \aa depending on centrality $b$. ${d\sigma_{dijet}/dy_{max}}$ is the pQCD dijet spectrum on parton rapidity $y_{max} = \ln(2E_{jet}/m_\pi)$. The FD shape is ensemble-mean FF $D_{\text{xx}}(y,b)$. Below fragment momentum 2 GeV/c the FD is dominated by the FF of 3 GeV jets.
$\epsilon_j(\Delta \eta) \in [1,2]$ measures the mean number of jets per dijet in acceptance $\Delta \eta$ (assuming point-like jets). $\Delta \eta_{4\pi}$ is  the effective $4\pi$ $\eta$ acceptance. Relating pQCD FDs to measured spectrum hard components provides constraints on parton spectrum parameters: the exponent of the power law and the effective spectrum cutoff energy which determines dijet total cross section $\sigma_{dijet}(b)$. 


The integral of $D_{xx}(y,y_{max})$ over fragment rapidity $y$ is dijet fragment multiplicity $2n_{ch,j}(y_{max})$.
Integrating both sides of Eq.~(\ref{fold}) over fragment rapidity gives the per-participant-pair fragment angular density on $\eta$
\bea \label{fold2}
\frac{dn_{h}}{d\eta}  \hspace{-.05in} &=&   \hspace{-.05in} \frac{\epsilon_j(\Delta \eta)}{\sigma_{NSD}\, \Delta \eta_{4\pi}}\int_0^\infty \hspace{-.1in}  dy_{max}\, n_{ch,j}(y_{max})\, \frac{d\sigma_{dijet}}{dy_{max}} \\ \nonumber
 &=&  \hspace{-.05in}  f(b)\,n_{ch,j}(b) = \frac{2\pi}{n_{bin}} J(b),
\eea
where $n_{ch,j}$ is the per-jet fragment multiplicity averaged over the minimum-bias parton spectrum, effectively the fragment multiplicity for partons near the 3 GeV spectrum cutoff.
The jet properties in Eq.~(\ref{fold2})---minimum-bias jet frequency $f(b)$ and mean jet fragment multiplicity $n_{ch,j}(b)$---relate jet angular correlations to spectrum hard components and pQCD fragment distributions~\cite{fragevo}. 



\subsection{Jet frequency from pQCD}

Jet frequency $f(b)$ 
is defined via Eq.~(\ref{fold2}) by
\bea \label{jetfreqq}
f(b) &\equiv&   \frac{1}{n_{bin}(b)} \frac{dn_{j}(b)}{d\eta}  \\ \nonumber
&=& \frac{\epsilon_{j}(b)\,\sigma_{dijet}(b)}{\sigma_{NSD} \Delta \eta_{4\pi}(b)},
\eea
where 
$\sigma_{NSD} = 36.5$ mb is the total cross section for non-single-diffractive (NSD) N-N collisions~\cite{nsd}, and $n_{bin}$ is the mean number of \nn binary collisions per \aa collision. The $f(b)$ estimate is constrained by comparisons between pQCD FD calculations and measured HCs~\cite{fragevo}. $\sigma_{dijet}(b)$ is determined by the parton spectrum cutoff inferred from hadron spectrum hard components.

In Fig.~\ref{j2b} (right panel) the number of jets per dijet within acceptance $\Delta \eta$ is $\epsilon_j(\Delta \eta) = 1/(1-a/2)$, with fractional $\eta$ acceptance $a = \Delta \eta / \Delta \eta_{4\pi}$. The points correspond to $a = 1/5, 2/5, 2/4$, three cases relevant to this analysis. The first two cases are for 200 GeV \pp collisions with $\Delta \eta = 1, 2$, the last for central \aa collisions with $\Delta \eta = 2$ and $\Delta \eta_{4\pi}$ reduced by 20\%. 

\begin{figure}[h]
\includegraphics[width=1.65in]{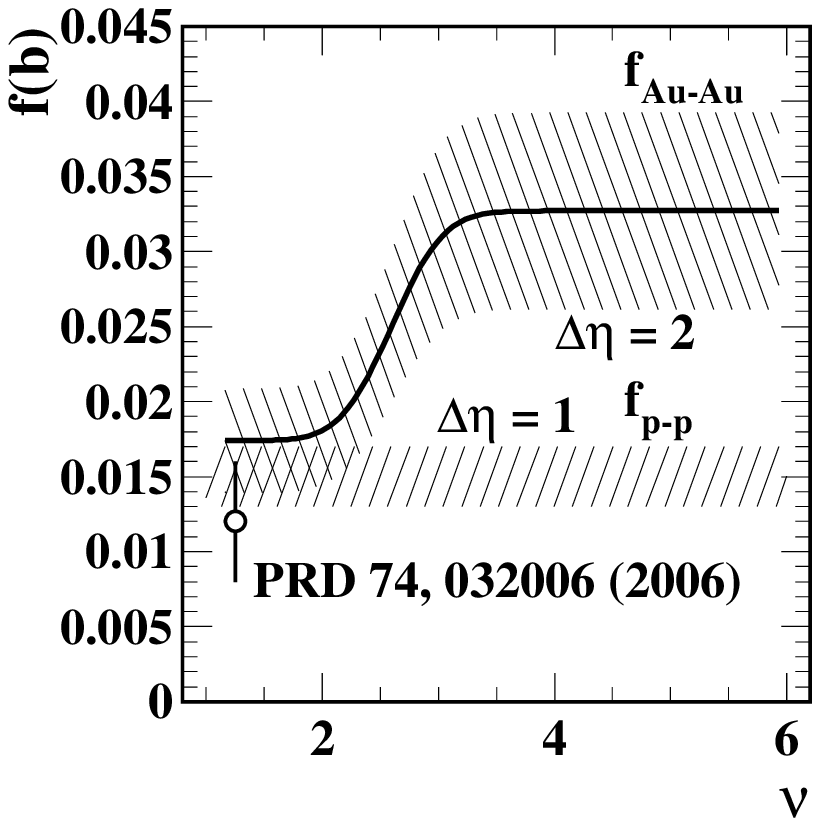}
\includegraphics[width=1.65in]{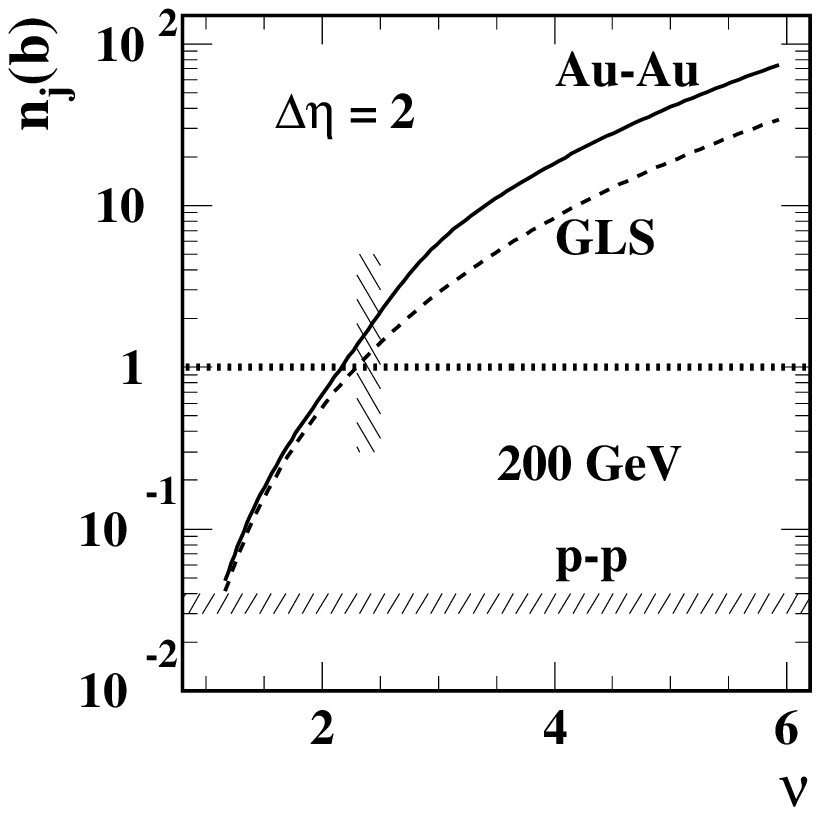}
\caption{\label{jetfreq} 
Left: Jet frequency $f(b) = (1/n_{bin})\, dn_j /d \eta$ vs centrality parameter $\nu$ for $\sqrt{s_{NN}} = 200$ GeV Au-Au collisions (upper hatched band) and for NSD \pp collisions (open circle~\cite{ppprd}, lower hatched band). Right: Corresponding total jet number $n_j(b)$ in angular acceptance $\Delta \eta = 2$.
} 
\end{figure}

In Fig.~\ref{jetfreq} (left panel) the solid curve describes the  estimated centrality variation of jet frequency $f(b)$ for 200 GeV \auau collisions in acceptance $\Delta \eta = 2$. The single data point is from Ref.~\cite{ppprd}, where an analysis of $p_t$ spectra from 200 GeV NSD \pp collisions within $\Delta \eta = 1$ invoked Eq.~(\ref{fold2}) (second line) with $H_{AA} = dn_h/2\pi d\eta$ and an estimate of $n_{ch,j}  = 2.5$ for $E_{jet} \sim 4$ GeV from Ref.~\cite{cdfmult} to determine $f_{pp} = (d n_h/d\eta) / n_{ch,j} = 0.03 / 2.5 = 0.012\pm 0.004$. 
%
The pQCD prediction for $f(b)$ from Eq.~(\ref{jetfreqq}) for \pp collisions within $\Delta \eta = 1$  is
\bea
f_{pp}   &=& \frac{1.1 \times \text{2.5 mb}}{\text{36.5 mb} \times 5} \\ \nonumber
&\approx& 0.015 \pm 0.0025,
\eea
consistent with Ref.~\cite{ppprd} and represented by the lower hatched band. For increased acceptance $\Delta \eta = 2$ $\epsilon_j \rightarrow 1.26$ from 1.1, so $f_{pp} \rightarrow 0.0175$ which fixes the left end of the solid curve, with 15\% uncertainty. 

In more-central \auau collisions a $1.6$-fold increase in the dijet cross section relative to \pp is inferred from comparisons of FDs to measured spectrum hard-component shapes~\cite{fragevo}. Retaining the effective $4\pi$ jet acceptance on $\eta$ assumed for \pp collisions the f(b) value is therefore 
\bea
f_{\text{Au-Au}}(b=0)   &=& \frac{1.26 \times \text{4 mb}}{\text{36.5 mb} \times 5} \\ \nonumber
&\approx& 0.028 \pm 0.0045.
\eea
If $\Delta \eta_{4\pi}(b)$ were to decrease by 20\% in central \auau collisions then $\epsilon_j$ increases to 1.35 and $f_{\text{Au-Au}}(\text{$b$=0}) \rightarrow 0.037$. The two values establish limits on the estimated $f(b)$ for central \auau collisions. The combined uncertainty for more-central collisions is then $\pm 20$\%. The transition from the \pp estimate to the range of values for more-central \auau collisions is assumed to be step-like based on observed sharp transitions in the $p_t$ spectrum hard component~\cite{hardspec} and jet angular correlations~\cite{daugherity}. The transition begins near $\nu = 2.5$ and ends near $\nu = 3.5$. See also Fig.~\ref{peakparams} (right panel).

Figure~\ref{jetfreq} (right panel) shows the number of jets $n_j(b) = n_{bin}\, \Delta \eta\,  f(b)$ within $\Delta \eta = 2$ for 200 GeV \auau collisions (solid curve) and the corresponding GLS reference (dashed curve) representing binary-collision scaling of \pp (N-N) collisions. The lower hatched band represents the value for \pp collisions. It is notable that deviations of jet properties from the GLS reference become significant where the jet number per unit $\eta$ becomes substantially greater than unity (upper hatched region) near $\nu \sim 2.5$, the location of the sharp transition in spectrum hard components and jet angular correlations.

\subsection{Fragment multiplicity from angular correlations} \label{jetmult}


We can now combine estimated jet frequencies with measured jet angular correlations to predict mean jet fragment multiplicities. The minimum-bias fragment multiplicity is obtained by introducing the $n_j(b)$  hypothesis into Eq.~(\ref{bigj2b}) to obtain
\bea \label{nchjj}
n_j(b)\, n_{ch,j}^2(b) &=&  (2\pi \Delta \eta)^2 \, J^2(b) \\ \nonumber 
&=& n_{ch}^2(b)\, j^2(b) ~~~\text{fragment pairs} \\ \nonumber
 n_{ch,j}(b) &=& n_{ch}(b)\, \sqrt{j^2(b) / n_j(b)},
\eea
where $n_{ch}(b) = 2\pi \Delta \eta\, \rho_0(b)$ (charged multiplicity in the angular acceptance) and $j^2(b)$ are measured quantities. 

\begin{figure}[h]
\includegraphics[width=1.65in]{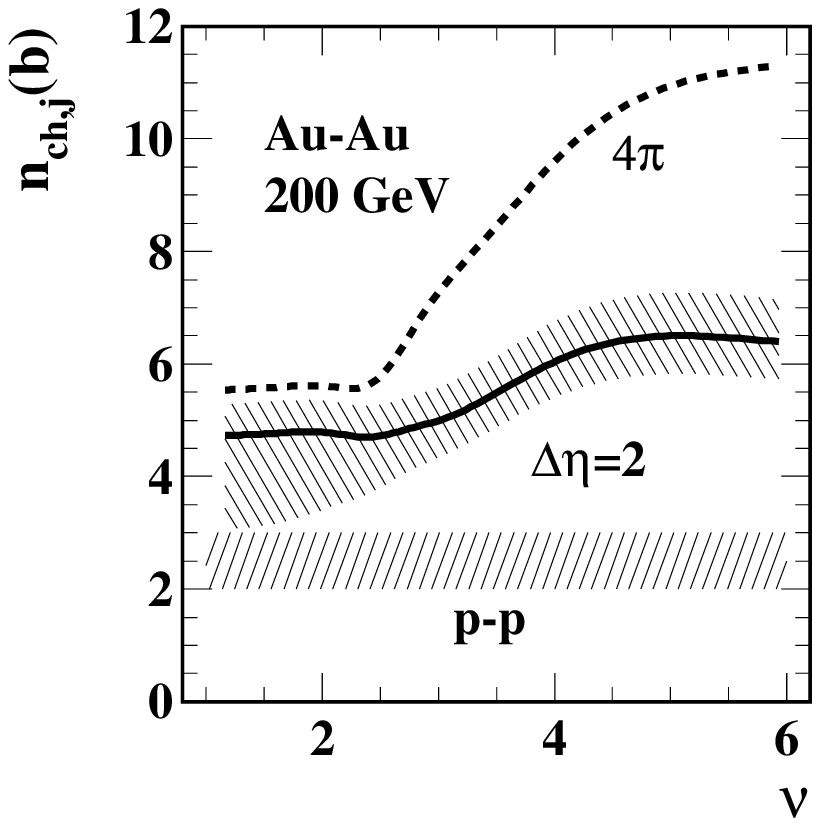}
\includegraphics[width=1.65in]{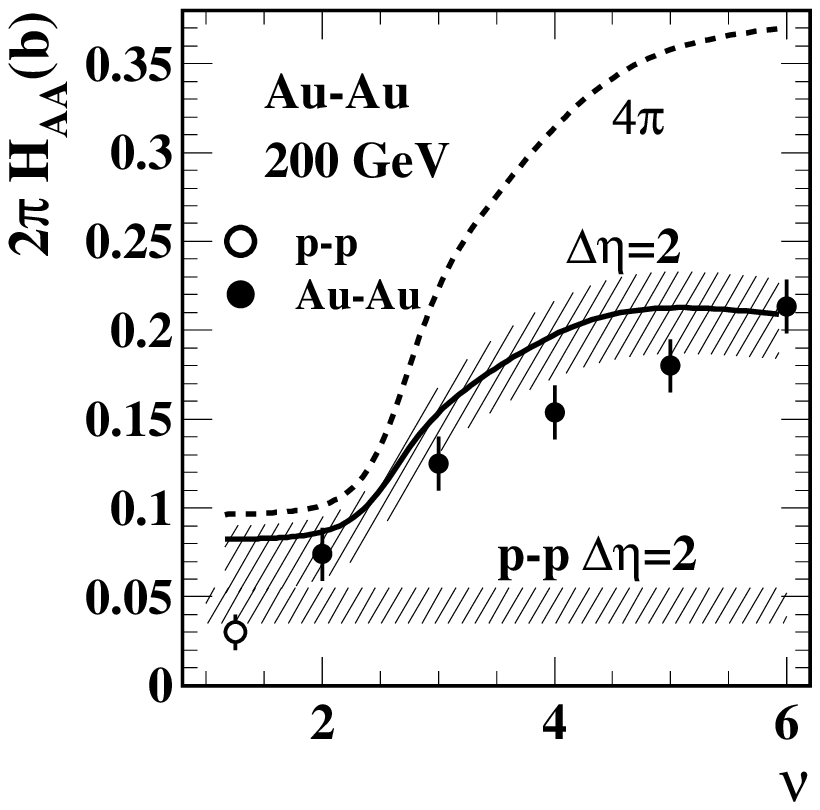}
\caption{\label{nchj} 
Left: Mean jet fragment multiplicity $n_{ch,j}(b)$ vs centrality parameter $\nu$  for $\sqrt{s_{NN}} = 200$ GeV \auau collisions (upper hatched band) and \pp collisions (lower hatched band). Right: Spectrum hard-component yield $2\pi\, H_{AA}(b)$ vs $\nu$ inferred from two-particle jet correlations (solid curve) and from single-particle spectra (solid points)~\cite{hardspec}. The lower hatched band and open point represent the hard-component yield obtained from NSD \pp collisions~\cite{ppprd}.
} 
\end{figure}

Figure~\ref{nchj} (left panel) shows rms jet multiplicity $n_{ch,j}(b)$ inferred from Eq.~(\ref{nchjj}) (third line) as a function of \auau centrality. The solid curve is the result within 2D angular acceptance $(\Delta \eta,\Delta \phi) = ( 2,2\pi)$ including edge losses. The dashed curve is  the result for a $4\pi$ acceptance with 100\% jet fragment efficiency.

While the detected jet fragment multiplicity for more-central \auau collisions is reasonably well determined, that for peripheral collisions appears to be overestimated when compared to single-particle results~\cite{fragevo}. From \pp spectrum hard-component measurements and from \ee and p-\=p FF systematics we expect \pp jet fragment multiplicity 2-3~\cite{cdfmult,ppprd,ffprd}. 
For minimum jet energy 3 GeV and mean fragment multiplicity 2-3 we obtain a most-probable fragment momentum $\sim$ 1.2 GeV/c,  consistent with observations~\cite{ppprd}. $n_{ch,j}\sim 4$-5 from jet correlations suggests a significant contribution to $j^2(b)$ from fluctuations, as discussed in App.~\ref{flucts}.

\section{Jet fragment angular density} \label{fragyield}

We now combine jet properties inferred from angular correlations to predict jet fragment densities which can be compared with measured spectrum hard components.

\subsection{Inferred spectrum hard component $\bf H_{AA}$}

Differential hard component $H_{AA}(y_t,b)$  represents the mean parton fragment spectrum per N-N binary collision within an A-A collision. For \aa transparency (GLS reference) $H_{AA}(b) = H_{NN}$ independent of centrality. $H_{AA}$ has been obtained from measured spectra by a subtraction procedure~\cite{ppprd,hardspec} and compared quantitatively to pQCD fragment distributions~\cite{fragevo}. {\em Integral} hard component $H_{AA}(b)$ represents by hypothesis the angular density of large-angle-scattered-parton fragments within an acceptance. In the present analysis we can infer $H_{AA}(b)$ from jet angular correlations by
\bea \label{haaeq}
2\pi  H_{AA}(b) &=& f(b)\, n_{ch,j}(b) \\ \nonumber
H_{AA}(b) &=& \frac{1}{n_{bin}}\, \rho_0(b)\, \sqrt{n_j(b)\, j^2(b)},
\eea
where the second line follows from Eq.~(\ref{nchjj}).

Figure~\ref{nchj} (right panel) shows $2\pi H_{AA}(b)$ obtained from jet angular correlations 
for limited $\eta$ acceptance and jet edge losses (solid curve) and for $4\pi$ acceptance with 100\%-efficient fragment detection (dashed curve). The \pp datum (open circle) was inferred from spectrum data with $\Delta \eta = 1$~\cite{ppprd}. The value  $2\pi H_{pp} =2.5 \times 0.012 =  0.03$ was based on a na\"ive Gaussian model for $H_{pp}(y_t)$ without exponential tail. For $\Delta \eta = 2$ and a more accurate model function we obtain 
\bea
2\pi\, H_{AA}(b) &\approx& 2.5\times 0.018  = 0.045,
\eea
defining a revised GLS reference with uncertainty marked by the lower hatched band.

The \auau data (solid points) are  derived from the ``total hadrons'' data in Fig.~15 (left panel) of Ref.~\cite{hardspec}. The consistency between jet correlations and spectrum hard components is  good, except for peripheral \aa collisions where there may be a significant fluctuation contribution to jet correlations, as discussed in App.~\ref{flucts}. A factor 4-5 increase in $H_{AA}$ with \auau centrality is suggested by spectrum and minimum-bias jet correlation data.

\subsection{Jet correlations and the two-component model}

\begin{figure}[h]
\includegraphics[width=1.65in]{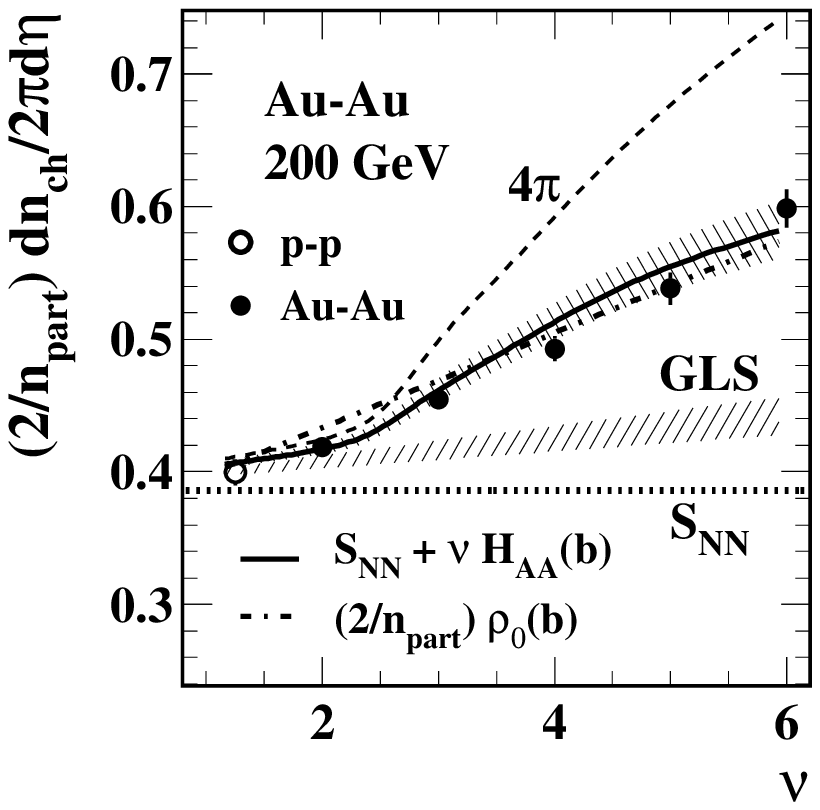}
\includegraphics[width=1.65in]{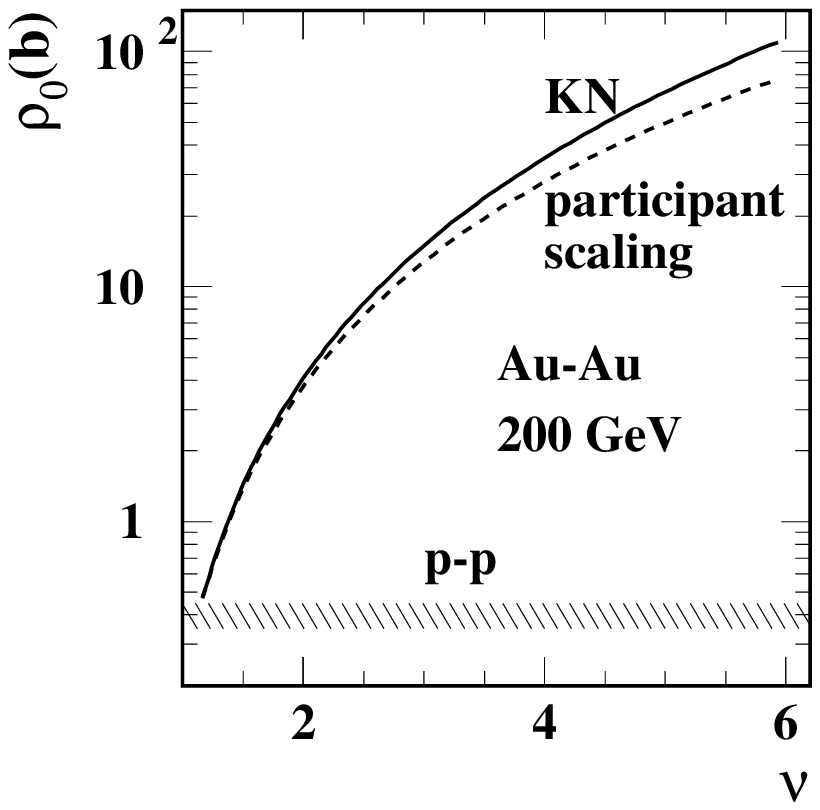}
\caption{\label{haa} 
Left: The per-participant-pair total hadron angular density $(2/n_{part})\, \rho_0(b)$ derived from two-particle jet correlations (solid curve) and from spectrum data (open~\cite{ppprd} and solid~\cite{hardspec} points). Note the suppressed zero. The lower hatched band represents a Glauber linear superposition (GLS) reference. Right: The corresponding Kharzeev-Nardi (KN) two-component model (solid curve) for the single-particle angular density (dash-dotted line in the left panel).
} 
\end{figure}

Multiplying Eq.~(\ref{haaeq}) through by $\nu / 2\pi$ or $\nu$ gives
\bea \label{nuhaa}
\nu H_{AA}(b) &=& \frac{2}{n_{part}}\, n_{ch,j}\frac{dn_j}{2\pi d\eta} \\ \nonumber
&=&  \frac{2}{n_{part}} \rho_0(b)  \sqrt{n_j(b)\, j^2(b)}.
\eea
$\nu H_{AA}(b)$ is the total hard component of the two-component spectrum model in  Eq.~(\ref{rho0b}) (first line).
Figure~\ref{haa} (left panel) shows the two-component particle yield $S_{NN} + \nu H_{AA}(b)$ predicted by {measured} jet angular correlations (bold solid curve). Soft component $S_{NN}$ is by hypothesis fixed at $\sim 0.4$ [2D density on $(\eta,\phi)$] for all  centralities. The solid points are the ``total hadrons'' data in Fig.~15 (left panel) of Ref.~\cite{hardspec} divided by $2\pi$. The dash-dotted line is the Kharzeev-Nardi (KN) approximation to per-participant charged-hadron 2D density $(2/n_{part}) \rho_0(b)$ assumed for this analysis~\cite{nardi}. 
$\nu H_{AA}$ increases by a factor 6$\times(4$-5) = 25-30 relative to the hard component in \pp collisions.

Figure~\ref{haa} (right panel) shows 2D angular density $\rho_0(b)$ on $(\eta,\phi)$. The solid curve is KN model $\rho_0(b) = (n_{part}/2) \rho_{pp}\left[1+ x(\nu - 1)\right]$ with $\rho_{pp} = 0.4$ and fixed $x = 0.09$ for Au-Au at 200 GeV~\cite{nardi}.  
The KN model describes minimum-bias data in more-central collisions but fails in more-peripheral collisions where corrected yield data are sparse. For more-peripheral collisions we expect a GLS trend extrapolated from p-p collisions with $x \sim 0.018$-0.035 (lower hatched band in the left panel). The sharp transition in jet angular correlations near $\nu = 2.5$ explains the large change in KN parameter $x$. 

According to Eq.~(\ref{nuhaa}) (second line) the {\em fractional hard component} $F \equiv \nu\, H_{AA} / \{(2/n_{part})\, \rho_0\} \approx 0.3$ from spectrum analysis of central \auau collisions~\cite{hardspec}, can also be obtained from angular correlations via $ \sqrt{n_j(b)\, j^2(b)}$. (F is the same quantity defined in~\cite{nardi}.) The first factor in the radicand is obtained from pQCD (relative systematic uncertainty $< 20$\%). The second factor is from measured jet angular correlations (relative uncertainty small). For central \auau collisions we obtain $\rho_0\, j^2 \sim 0.165$, $\rho_0 \sim 110$ and $n_j \sim 70$, leading to $F = 0.32$. That result is consistent with the analysis in Ref.~\cite{nardi}, but in this case inferred directly from jet correlations.

We thus conclude that the observed increase in particle production beyond participant scaling in central \auau collisions is fully explained in terms of jet angular correlations. 
Combining angular correlation measurements and a pQCD estimate of jet number we find that {\em one third of the final state} in 200 GeV central Au-Au collisions is associated with {\em resolved jet correlations} (relative uncertainty $< 10$\%).


\section{Systematic uncertainties}

Systematic uncertainties for jet-related particle production have different trends below and above the ``sharp transition'' in jet characteristics near $\nu = 2.5$. Interpretation of fragment yields from correlations for more-peripheral collisions is less certain due to fluctuation contributions but those results can be augmented by single-particle studies of elementary collisions (p-p, $e^+$-$e^-$) and by pQCD Monte Carlo results. Systematic uncertainties from fluctuations for more-central \aa collisions are smaller due to  increased fragment multiplicities. 

\subsection{Angular correlation measurements}

The primary data source for this analysis is pair ratio $j^2(b)$ derived from fits to the same-side 2D peak and averaged over the angular acceptance. The numerical uncertainty in the underlying 2D histograms is negligible compared to other uncertainties in the analysis. Derivation of $j^2(b)$ from 2D histograms involves model fits which, for the same-side 2D peak, have uncertainties of order 5\% since the 2D peak is a dominant correlation structure.

\subsection{Inferred jet properties}

Uncertainty in mean event-wise jet number $n_j(b) \leftrightarrow f(b)$ is substantial due to uncertainty in $\sigma_{dijet}(b)$ and $\Delta \eta_{4\pi}$, as shown in Fig.~\ref{jetfreq} (left panel). $\sigma_{dijet}(b)$ depends on comparisons between calculate pQCD fragment distributions and spectrum hard components in \pp and \auau collisions, specifically the location of the mode of the hard component on $y_t$~\cite{fragevo}. For \pp collisions the cross-section estimate was 2.5$\pm 0.6$ mb, or a 25\% relative uncertainty. Evolution of the spectrum hard component with centrality led to inference of a $4\pm1$ mb cross section in central \auau due to a 10\% downward shift of the effective cutoff energy. We also include the possibility that the effective $4\pi$ $\eta$ acceptance may be reduced in more-central \auau collisions. The overall uncertainties are summarized by the hatched regions in Fig.~\ref{jetfreq} (left panel).

Uncertainties in mean jet fragment multiplicity $n_{ch,j}(b)$ are indicated by the hatched regions in Fig.~\ref{nchj} (left panel). The upper hatched region describes the multiplicity estimate derived from angular correlations assuming pair factorization. Uncertainties for more-central \aa collisions are dominated by the 20\% contribution from the jet frequency. However, the square root reduces the relative uncertainty to approximately 10\%. In more-peripheral collisions possible fluctuation contributions to $j^2(b)$ substantially increase the uncertainty in  $n_{ch,j}(b)$ (see App.~\ref{flucts}). However, information from elementary collisions can be invoked to supplement the multiplicity estimate there. The lower hatched region describes a fragment multiplicity estimate based on measured fragmentation functions and their uncertainties. 


\subsection{Jet fragment yields and hadron production}

Jet fragment yields can be estimated by $2\pi\, H_{AA}(b) = f(b)\, n_{ch,j}(b)$. However, because the uncertainties in the factors are strongly correlated they do not add quadratically. In fact, the relative uncertainties in $H_{AA}(b)$ and $n_{ch,j}(b)$ are the same, since $n_{ch,j}(b) \propto \sqrt{j^2(b) / n_j(b)}$ and
$ H_{AA} \propto \sqrt{j^2(b)\, n_j(b)}$. Factors $n_{ch}(b)$ and Glauber parameters omitted in that comparison have relatively small uncertainties. The factors in the radicands have independent uncertainties, with the 20\% for $n_j$ dominating. Because of the square root the relative uncertainty in both quantities is about $\pm 10$\% for more-central collisions. For peripheral collisions the uncertainty increases because of the fluctuation contribution.


The per-participant-pair total charged-particle yield in the two-component model is $(2/n_{part})\, \rho_0(b) = S_{NN} + \nu H_{AA}(b)$. 
The relative uncertainty in the {\em total} hadron yield as inferred from jet correlations is significant for central collisions but negligible for peripheral collisions, since $\nu H_{AA}$ increases by a factor 30 with centrality. 






\section{Discussion}

The direct comparison in Figs.~\ref{nchj} (right panel) and \ref{haa} (left panel) between previously-measured spectrum hard components and comparable data inferred from a jet-like feature in angular correlations (present analysis) seems to provide substantial additional support for a minimum-bias jet interpretation. Independent procedures based on a jet hypothesis agree within estimated uncertainties and with a pQCD calculation.
%

\subsection{Do pQCD jets contribute to low-$p_t$ structure?}

According to pQCD theory jets (correlated hadron fragments of scattered partons) should contribute significant structure in both correlations and single-particle spectra. Jets should appear in spectra as a hard-component contribution over some $p_t$ interval. Jets should also appear as a same-side 2D peak and away-side ridge in combinatoric two-particle correlations. But, how far down in $p_t$ does ``true'' jet structure extend? Is a jet description valid for minimum-bias jets with energy peaked near 3 GeV and fragments with $p_t\sim 1$ GeV/c?

Phenomenologically, spectrum hard component $H_{AA}$ is defined for more-peripheral A-A collisions in the two-component model as the part of the spectrum that scales as $n_{bin}$, or as $\nu$ relative to $n_{part}/2$ (participant scaling)~\cite{hardspec}. That definition emerged from a physical-model-independent analysis of \pp collisions~\cite{ppprd}. Later comparisons to pQCD calculations lent support to a jet fragment interpretation down to $\sim 0.3$ GeV/c~\cite{fragevo}.

In the present study jet-like structure in $p_t$-integral two-particle correlations~\cite{daugherity} is compared with $H_{AA}(y_t,b)$ from spectra integrated to obtain yields $H_{AA}(b)$~\cite{hardspec}. The detailed analysis assumes a jet mechanism for correlation structure and the validity of pQCD applied to that structure. The agreement is remarkable. Since the mode of jet-correlated particles is equivalent to $p_t = 1$ GeV/c~\cite{porter1,porter2} we conclude that true jet correlations must extend significantly below 1 GeV/c. The quantitative agreement of spectrum structure, correlation structure and pQCD calculations supports a consistent minimum-bias jet picture: Correlated fragments from minimum-bias partons ($\sim 3$ GeV) play a major role in all RHIC collisions down to about 0.3 GeV/c.

\subsection{Plotting formats and fragmentation}

References~\cite{nohydro,tomaustin} provide several examples of the consequences of format choices. Some conventional plotting formats effectively minimize manifestations of parton fragmentation. A comparison between Fig.~\ref{haa} (left panel) and Fig.~\ref{nchj} (right panel)  provides an illustration. 

In Fig.~\ref{haa} (left panel) both spectrum data (points) and correlations (solid curve) seem to be consistent with the linear KN model (dash-dotted line). Even with the substantial plot offset it is difficult to see any effect of the sharp transition, and it could be argued from that plot that there is none. 
In  Fig.~\ref{nchj} (right panel) the more differential format clearly shows significant deviations, since the  KN model in that format would correspond to a constant value proportional to the fixed KN $x$ parameter (see App.~\ref{knn}). Correlation and spectrum data provide clear evidence for the sharp transition in jet characteristics near $\nu = 2.5$.

Comparing the structure in Fig.~\ref{peakparams} and the examples above we observe a progression from correlated pairs (transition dominating) to differential hard-component particle yields (transition clearly apparent) to total hadron yields (transition effectively concealed).
 
\subsection{New access to low-energy jet systematics}

Factorization of minimum-bias jet correlations provides new access to pQCD processes at low parton energies and fragment momenta where most of the fragment yield appears in nuclear collisions. The SS 2D peak, interpreted in terms of {minimum-bias} jets, is quantitatively connected to pQCD through pair factorization and the spectrum hard component. The connection persists down to parton energy 3 GeV and hadron momentum zero. 

Jet fragment yield $n_{ch,j}(b)$ is effectively the fragment multiplicity of 3 GeV jets because of the parton power-law spectrum cutoff. Fragment yields inferred from correlations can be compared with 3 GeV multiplicities extrapolated from \ee FFs~\cite{ffprd} and from \ppbar jets~\cite{fragevo}, with and without proposed medium-modification effects. 


Inferred jet frequence $n_j(b)$ is of central importance to RHIC physics. It describes the number of jets that appear as correlated fragments in the final state, integrated over the entire parton spectrum.  Minimum-bias correlation data establish a constraint on $n_j(b)$ that agrees with the spectrum hard-component analysis. Those results imply that essentially all initial-state large-angle-scattered partons down to 3 GeV observed in \pp collisions also survive to the final state in central \auau collisions (scaled per binary \nn collision) as resolved jet angular correlations.

\subsection{Implications for RHIC collisions}

The study of QCD dynamics at RHIC has competed with a strong emphasis on hydro models and possible QGP formation. The importance of fragmentation in RHIC collisions has become more apparent in several recent studies. An emerging issue for \aa collisions is the major changes in parton scattering and fragmentation that occur at smaller $p_t$ within larger space-time volumes. 


Attempts to measure radial flow with identified-hadron spectra in  the context of a two-component spectrum model failed~\cite{hardspec}. Against expectations all spectrum evolution with centrality was confined to the hard component, first isolated in \pp spectra~\cite{ppprd}. The spectrum hard component was then quantitatively related to pQCD calculations~\cite{fragevo}, supporting the conclusion that the hard component represents minimum-bias parton fragmentation.

Hydro interpretation of the azimuth quadrupole structure as ``elliptic flow'' is also questionable. Strong jet (``nonflow'') contributions to published $v_2$ data have confused their interpretation~\cite{gluequad}. Recent measurements of $v_2$ with 2D angular correlations which eliminate jet contributions reveal systematics trends inconsistent with hydro~\cite{kettler}. Reassessment of published $v_2$ data in a broader context suggests an alternative interpretation in terms of novel QCD field phenomena~\cite{gluequad,quadspec}.


The present analysis strongly suggests that a substantial fraction of the final state in central Au-Au collisions consists of resolved jets with energies as low as 3 GeV. Parton scattering and fragmentation provide a common mechanism for both spectrum hard components and jet-like correlations down to small parton energies and hadron fragment momenta. The evolution of nuclear collisions is apparently dominated by parton fragmentation even in the most central \auau collisions, albeit fragmentation is strongly modified there. Those results pose significant difficulties for the RHIC ``perfect liquid'' paradigm. Several examples follow:


{\bf Multiple scattering of partons and hadrons --} Multiple scattering might lead to formation of thermalized partonic and/or hadronic media. 

{\bf Response --} In the present analysis we find no loss of initial scattered partons to thermalization, only redistribution of parton energy {\em within jets} during fragmentation. Strong jet correlations persist for low-energy partons and low-momentum hadrons, contradicting significant multiple scattering of either partons or hadrons. In particular, the same-side 2D peak {\em narrows} on azimuth, which is inconsistent with parton or hadron multiple scattering.

{\bf Formation of a thermalized ``opaque core'' --} Parton multiple scattering may establish a dense thermalized medium (opaque core) which absorbs most partons. Any surviving jets are emitted from a ``corona'' region at the (radial) surface of the collision system. Imposition of high-$p_t$ triggers biases toward ``tangential emission'' from the corona and unmodified jet pairs.

{\bf Response --} Focusing only on yield reductions in high-$p_t$ bins ($R_{AA}$, ``triggered'' jet correlations) to conclude parton absorption in a dense medium is misleading. Reduced fragment number at larger $p_t$ should not be interpreted as a reduction in the precursor-parton number. The error in inference may be large due to the steepness of the fragment spectrum arising from the underlying parton power-law spectrum.  Final-state jet structure should be studied over the full hadron $p_t$ acceptance to best understand parton precursors. The present analysis combined with other studies indicates that the number of {\em resolved} final-state jets per \nn collision {\em increases} with \aa centrality. Fragmentation is modified, with more fragments at smaller momentum as the general trend.

{\bf Distortion of the ``away-side'' jet implies ``Mach cones'' --} Energetic partons passing through a dense medium may produce Mach shocks manifested as a double-peaked structure in the ``away-side'' jet. Such distortions (azimuth double peak near $\pi$) are interpreted to imply formation of Mach shocks in the dense medium.

{\bf Response --} All jets within the angular acceptance must appear in the {\em minimum-bias} same-side (SS) jet peak. Any deformation of ``away-side'' jets should then appear in the SS peak. But that is not observed. The ``away-side'' jet is a fiction. The away-side peak at $\pi$ radians reflects {\em inter}jet correlations {\em between} jets.  Broadening of the away-side peak reflects  acoplanarity of the parton partners (e.g.\ $k_t$ broadening), not internal jet structure. Even the lowest-energy jets are not deformed azimuthally~\cite{daugherity}. The double-peaked structure attributed to Mach cones arises from $v_2$ data which may include a substantial jet contribution (``nonflow'')~\cite{tzyam}.

{\bf Thermalized low-energy scattered partons drive hydro flows --} Scattered partons at energy scales of a few GeV may undergo multiple scattering to form a dense thermalized partonic medium. The large thermalized parton flux leads to large energy densities and pressure gradients which then drive hydrodynamic flows~\cite{hydrotheory}.

{\bf Response --} The present analysis combined with two-component spectrum analysis~\cite{hardspec} and direct comparisons of spectrum data with pQCD calculations~\cite{fragevo} indicates that {all} initial-state large-angle-scattered partons appearing as jets in \pp collisions (down to a 3 GeV parton spectrum cutoff) also appear as resolved jets in central \auau collisions. None of the expected pQCD parton spectrum is lost to thermalization. Partons do not contribute to large energy densities or pressure gradients. These conclusions are consistent with failure to observe significant radial flow in differential spectrum analysis of 200 GeV \auau collisions with identified hadrons~\cite{hardspec}.

\section{Summary}

Minimum-bias jet (minijet) angular correlations have been converted to absolute parton fragment yields which are found to comprise approximately one third of the hadronic final state in 200 GeV central Au-Au collisions. Direct comparison of minijet correlations with previously-measured spectrum hard-component yields reveals good agreement within data uncertainties.  


pQCD fragment distributions  calculated by folding a minimum-bias parton energy spectrum with a parametrization of measured fragmentation functions accurately describe measured spectrum hard components. The combined results reveal that almost all large-angle scattered partons down to 3 GeV parton energy survive as true jet manifestations in spectra and correlations, albeit with significant modification of fragmentation.

Large jet contributions to spectra and correlations quantitatively  described by pQCD theory contradict claims of early thermalization by parton multiple scattering and formation of a strongly-coupled, small-viscosity QGP. Certainly we observe significant modification of parton fragmentation, but the basic pQCD processes persist even in the most-central \auau collisions at RHIC.

Hydro-motivated analysis of RHIC data tends to interpret the large parton fragment contribution below 2 GeV/c in terms of flow phenomena. pQCD descriptions are artificially restricted to small regions of momentum space. The role of parton fragmentation is thereby minimized. Model-independent analysis of spectrum and correlation structure as in the present study reveals new fragmentation features quantitatively described by pQCD over the full hadron momentum range.

\begin{appendix}

\section{Glossary and Symbols}

\begin{quote}

\begin{description}
\itemsep .0in
\itemindent -.4in

\item[Fragmentation function:] (FF) Parton fragment spectrum conditional on parton energy.

\item[Minimum-bias parton spectrum:] Energy spectrum of all large-angle-scattered partons fragmenting to (charged) hadrons (jets at midrapidity). A parton spectrum cutoff near 3 GeV is observed in \pp collisions.


\item[pQCD fragment distribution:] (FD)  Conditional parton fragment spectra (FFs) folded with a minimum-bias parton spectrum.

\item[Minimum-bias parton fragments:] All (charged) hadrons within the angular acceptance from fragmentation of minimum-bias partons.

\item[Minimum-bias jets (minijets):] Jet fragment angular correlations from a minimum-bias parton spectrum, mainly 3 GeV jets.

\item[Two-component model:]  Spectra and correlations separated into soft and hard components.

\item[Soft component:] (SC) Particles and correlated pairs from longitudinal fragmentation of projectile nucleons (soft Pomeron exchange).

\item[Hard component:] (HC) Particles and correlated pairs from transverse fragmentation of large-angle-scattered partons (hard Pomeron exchange).

\item[2D angular autocorrelation:] Projection from 4D pair angle space onto 2D difference variables $\eta_1 - \eta_2$ and $\phi_1 - \phi_2$ near mid-rapidity which retains all angular correlation information.

\item[Jet-like correlations:] Structure in 2D angular correlations on $\eta$ and $\phi$ difference variables including a same-side 2D peak at the origin and an away-side ``ridge'' uniform on $\eta$ difference.

\item[Glauber linear superposition reference:]  (GLS) A model of \aa collisions as linear superpositions of \nn collisions based on the two-component model. Deviations from the GLS indicate novel physics in \aa collisions.

\hspace{.0in}

\item[$\bf n_j(b)$:]  Minimum-bias jet number per \aa collision

\item[$\bf f(b)$:] Minimum-bias jet frequency per \nn binary collision, $= (1/n_{bin})\, dn_j/d\eta$.

\item[$\bf n_{ch,j}(b)$:] Per-jet mean fragment multiplicity.

\item[$\bf J^2(b)$:] Jet fragment pair 4D angular density.

\item[$\bf  J(b)$:] Jet fragment 2D angular density on ($\eta,\phi$).

\item[$\bf dn_h/d\eta$:]  Jet fragment 1D angular density on $\eta$.

\item[$\bf S_{NN}$:] Spectrum soft component, 2D angular density.

\item[$\bf H_{AA}(b)$:] Spectrum hard component, 2D angular density, $= dn_h/2\pi d\eta$.

\item[$\bf n_s$:]  Soft-component hadron multiplicity in $\Delta \eta$.

\item[$\bf n_h(b)$:]  Hard-component hadron multiplicity in $\Delta \eta$.

\item[$\bf n_{ch}(b)$:]  Total charged multiplicity in $\Delta \eta$, $ = n_s + n_h$.

\item[$\bf \rho_0(b)$:] Charged-particle 2D angular density, $= dn_{ch} / 2\pi d\eta \approx n_{ch}/2\pi \Delta \eta$.

\item[$\bf n_{part}/2$:] Mean number of participant projectile-nucleon pairs per \aa collision.

\item[$\bf n_{bin}$:] Mean number of \nn binary collisions per \aa collision.

\item[$\bf \nu$:] centrality measure $2 n_{bin} / n_{part}$, mean projectile-nucleon path length in number of encountered nucleons, $\in [1,6]$ for \auau collisions.

\end{description}
\end{quote}


\section{2D angle averages} \label{angav}


The starting point of the present analysis is a set of histograms $\Delta \rho / \rho_{ref}$ or $\Delta \rho / \sqrt{\rho_{ref}}$ on $(\eta_\Delta,\phi_\Delta)$ for some conditions on $(y_t,y_t)$ (joint or marginal distribution on $y_t$ or $y_t$-integral) and centrality $b$. Model fits are used to extract pair ratio $ j^2(\eta_\Delta,\phi_\Delta,b)$ describing the SS 2D peak according to the second line of Eq.~(\ref{estruct}).
Fit parameters $A_{2D}$, $\sigma_{\eta}$ and $\sigma_{\phi}$ may depend on some or all of conditions $(y_{t1},y_{t2},b)$. The model function may extend beyond the $\eta$ acceptance $\Delta \eta$. 
%
The {mean} pair ratio from the SS 2D peak {\em within} angular acceptance $(\Delta \eta,2\pi)$ is then
\begin{widetext}
\bea \label{average}
j^2(b) &=&    
 \hspace{-.0in} \frac{\int_{-\Delta \eta}^{\Delta \eta}\int_{-\pi}^{\pi} d\eta_\Delta   d\phi_\Delta \left(1-\frac{\eta_\Delta}{\Delta \eta}\right) 
j^2(b,\eta_\Delta,\phi_\Delta)}
{\int_{-\Delta \eta}^{\Delta \eta}\int_{-\pi}^{\pi} d\eta_\Delta  d\phi_\Delta\left(1-\frac{\eta_\Delta}{\Delta \eta}\right) = 2\pi\,\Delta \eta }   \\ \nonumber
 &=& A_{2D}\, \frac{\sqrt{2\pi \sigma^2_{\eta}}}{\Delta \eta}\frac{\sqrt{2\pi \sigma^2_{\phi}}}{2\pi} 
 \hspace{-0in} \times \left\{  {\rm erf} \left(\frac{\Delta \eta}{\sqrt{2\sigma^2_{\eta}}}\right) -  \sqrt{\frac{2 \sigma^2_{\eta}}{\pi\, \Delta \eta^2}}
\left[  1-\exp\left(-\Delta \eta^2 / 2 \sigma^2_{\eta}\right) \right]\right\} \\ \nonumber
&\equiv&A_{2D} \frac{\sigma_\eta\sigma_\phi}{\Delta \eta}\, {\cal A}(\sigma_\eta/\Delta \eta),
\eea
\end{widetext}
with pair acceptance represented by factor $\left(1-{\eta_\Delta}/{\Delta \eta}\right)$.  
Pair acceptance factor ${\cal A}$ (quantity in curly brackets) has limiting values ${\cal A}(\sigma_{\eta},\Delta \eta) \rightarrow 1$ as $\sigma_{\eta}/\Delta \eta \rightarrow 0$ and ${\cal A}(\sigma_{\eta},\Delta \eta) \rightarrow \Delta \eta /\sqrt{ 2\pi \sigma^2_{\eta}}$ for $\sigma_{\eta}/\Delta \eta \rightarrow \infty$. For a very wide peak on $\eta_\Delta$ the limiting value is $j^2 = A_{2D}\, \sigma_{\phi}/\sqrt{ 2\pi}$.

\section{Fluctuation effects}  \label{flucts}

Fluctuations in total charged-particle number, jet number per event and fragment number per jet may bias inferred jet systematics. The primary random variable is measured pair ratio $j^2(b)$, with ensemble-mean value 
\bea
j^2(b) &\equiv& \overline{\left\{\frac{n_j\, n^2_{ch,j}}{n_{ch}^2}\right\}}  \approx \frac{1}{\bar n^2_{ch}}\, \overline{n_j\, n^2_{ch,j} }
\eea
The RHS approximation is justified because $j^2(b)$ mean values are obtained within small bins on $n_{ch}$ to minimize systematic errors in the correlation analysis~\cite{daugherity}. The second factor is then decomposed into primary mean values, variance difference and covariance by
\bea \label{njnchj}
\overline{n_j\, n^2_{ch,j}} &=& \bar n_j \, \left\{\bar n_{ch,j}^2 + \bar n_{ch,j} + \Delta\sigma^2_{n_{ch,j}}\right\} + \sigma^2_{n_j,n^2_{ch,j}},
\eea
where $\Delta\sigma^2_{n_{ch,j}} = \sigma^2_{n_{ch,j}} - \bar n_{ch,j}$ is the excess variance relative to a Poisson reference, and $\sigma^2_{n_j,n^2_{ch,j}}$ is the covariance between the two quantities in its subscript. The assumption in Eq.~(\ref{jb}) (third line) is complete factorization of the LHS of Eq.~(\ref{njnchj}). Correction factor $G(b)$ for inferred $\bar n_{ch,j}(b)$ is defined by the ratio
\bea
\frac{\overline{n_j\, n^2_{ch,j}}}{ \bar n_j\,\bar n_{ch,j}^2} \hspace{-.05in} &=& \hspace{-.05in} 1  \hspace{-.03in}+ \hspace{-.03in} \frac{1}{\bar n_{ch,j}}\left\{1 \hspace{-.03in}+ \hspace{-.03in} \frac{\Delta\sigma^2_{n_{ch,j}}}{\bar n_{ch,j}} \hspace{-.03in} + \hspace{-.03in} \frac{1}{\sqrt{\bar n_j}} \frac{\sigma^2_{n_j,n^2_{ch,j}}}{\sqrt{\bar n_j \bar n^2_{ch,j}}}\right\}  \\ \nonumber
&=& 1 +\frac{1}{\bar n_{ch,j}} \left\{1+ A(b) + \frac{1}{\sqrt{\bar n_j}}\, B(b) \right \} \equiv G^2(b),
\eea
where $A(b)$ is a normalized (scaled) variance difference~\cite{inverse} and $B(b) \in [-1,1]$ has the form of Pearson's normalized covariance~\cite{pearson}.
 Covariance $\sigma^2_{n_j,n^2_{ch,j}}$ is proportional to the slope of the $n_{ch,j}$ trend in Fig. 4 (left panel). The maximum covariance then occurs near the sharp transition. 
Generally, fluctuations lead to overestimation of jet fragment mean multiplicities inferred from jet correlations, but larger fragment multiplicities in more-central \aa collisions reduce the bias.

As an example, suppose  low-energy gluons fragment only to several charge-neutral pion pairs and the relative frequency on multiplicities 2, 4, 6 is 4:2:1. Then $\bar n \approx 3$, $\sigma^2_n = 9$, $\Delta \sigma^2_n = 6$ and $A = 2$. Neglecting the covariance term $G^2 = 1 + \frac{1}{3}(1 + 2) = 2$. If the fragment multiplicity inferred from angular correlations by na\"ive pair factorization is 4 the true multiplicity is $4 / G \sim 2.8$.

\section{The KN model} \label{knn}


The Kharzeev-Nardi (KN) two-component model of \aa particle production is described by~\cite{nardi}:
\bea \label{kneq}
 \frac{2}{n_{part}}\,  \rho_0(b)  
&=&   \rho_{pp}\, \left\{1 + x\,(\nu-1) \right\}.
\eea
In the KN model the hard-component parameter is a fixed quantity $x \sim 0.1$ for more-central 200 GeV Au-Au collisions. However, $x$ for \pp and peripheral \aa collisions is smaller. If \aa collisions were linear superpositions of \nn collisions (GLS) we should expect a significantly smaller particle yield in central \aa collisions . 



Starting with the total per-participant-pair 2D angular density we can infer a centrality-dependent $x(b)$
\bea \label{2compform}
\frac{2}{n_{part}}\, \rho_0(b) &=&  S_{NN} + \nu\, H_{AA}(b) \\ \nonumber
&=& \rho_{pp}+\nu\, H_{AA}(b) -  H_{NN} \\ \nonumber
&=& \rho_{pp} \left\{  1 + \frac{H_{AA}(b)}{\rho_{pp}}\left[\nu  - 1/r_{AA}(b)\right]   \right\} \\ \nonumber
&=& \rho_{pp} \{ 1 + x(\nu - 1) \},
\eea
where $r_{AA}(b) = H_{AA}(b) / H_{NN}$~\cite{hardspec}.
We then identify $x(b) \approx {H_{AA}}(b)/{\rho_{pp}}$. Replacement of 1 by $1/r_{AA}(b)$ in KN factor $(\nu - 1)$, with $r_{AA}(b) \rightarrow 10$ in central collisions (increase of soft fragments~\cite{fragevo}), implies that the effective $x(b)$ further increases by factor $\nu /(\nu - 1) \sim 1.2$ in central collisions with $\nu \sim 6$. That is comparable to systematic uncertainties. $H_{AA}(b)$ in Fig.~\ref{haa} (left panel) then indicates the functional form of $x(b)$.

Since $\rho_{pp} \approx 0.4$ we obtain $x(b) \approx H_{AA}(b) / (0.4\times 2\pi)$ = $0.4H_{AA}(b)$. Specific values are $x_{pp}  \sim 0.02$ and $x(b=0)  \sim 0.09$ based on $H_{AA}(b)$ from spectrum hard components (\pp and peripheral Au-Au) and minijet correlations from this analysis (central Au-Au). $x(b)$ thus increases four- to five-fold from \pp to central \auau collisions, with a nearly-constant value above 50\%-central collisions.



\end{appendix}


\end{document}